\documentclass[12pt]{article}
\usepackage[dvips]{graphicx}

\begin{document}

\title{Widom-Rowlinson model (continuum and lattice)}

\author{Ladislav {\v S}amaj\footnote{Institute of Physics, 
Slovak Academy of Sciences, D\'ubravsk\'a cesta 9,
845 11 Bratislava, Slovakia; E-mail: Ladislav.Samaj@savba.sk}}

\date{\empty}

\maketitle

\noindent {\bf Continuum Widom-Rowlinson model:}
In the beginning of seventies, a great progress has already been
made in understanding equilibrium phase transitions and 
critical phenomena in classical (i.e. non-quantum) 
lattice models, where basic elements are localized 
on vertices of a regular discrete structure and interact with 
their nearest neighbors.
The lattice gas analogy \cite{Lee} of Onsager's exact solution 
of the two-dimensional Ising model enabled one to understand 
qualitatively as well as quantitatively physical implications of 
the spontaneous breaking of the particle-hole symmetry on the existence 
and critical properties of high-density liquid and low-density vapor phases.
On the other hand, little was known about the liquid-vapor equilibrium 
in continuum models (fluids), where the distance between 
interacting particles is a continuous variable.
In 1970 Widom and Rowlinson (WR) introduced a simple model
of the classical fluid in thermodynamic equilibrium \cite{Widom},
found a counterpart of the particle-hole symmetry and described 
implications of this symmetry on a potential liquid-vapor phase diagram. 
Subsequent studies proved rigorously the existence of 
liquid-vapor phase transitions in the WR and related models 
in spatial dimensions $d\ge 2$; 
up to now WR fluids are the only continuum systems with decaying 
interactions for which on has this kind of rigorous results.

The WR model consists of identical molecules living in an infinite
$d$-dimensional continuous space of points 
${\bf r}\in V$ $(V\to {\rm R}^d)$, around the center of each molecule 
there is a microscopic sphere of radius $\sigma$ and volume $v_0$.
The potential energy $U$ associated with a given configuration 
${\bf r}_1, \ldots, {\bf r}_N$ of $N$ molecules is defined by
\begin{equation} \label{1}
U({\bf r}_1, \ldots, {\bf r}_N) = \left[ W({\bf r}_1, \ldots, {\bf r}_N) 
- N v_0 \right] \epsilon / v_0 ,
\end{equation}
where $W$ denotes the volume covered by the corresponding $N$ (in general
interpenetrating) spheres and $\epsilon>0$ is some energy constant. 
Since $W$ fulfills the evident inequalities 
$v_0\le W({\bf r}_1, \ldots, {\bf r}_N) \le N v_0$, 
the potential energy is bounded as follows
\begin{equation} \label{2}
- (N-1) \epsilon \le U({\bf r}_1, \ldots, {\bf r}_N)  \le 0 ;
\end{equation}
the lower bound ensures a correct thermodynamics, the upper bound tells us 
that the short-ranged forces among molecules are purely attractive.
The model is usually studied in the grand canonical formalism 
characterized by the dimensionless inverse temperature
$\theta=\epsilon/(k_{\rm B}T)$ and particle activity $z$,
the corresponding particle density $\langle N\rangle/V$ and
pressure $P$ are considered in the dimensionless combinations
\begin{equation} \label{3}
\rho(z,\theta) = v_0 \langle N\rangle/V , \quad
{\cal P}(z,\theta) = P v_0/(k_{\rm B}T) ;
\end{equation}
$z$ is normalized so as to be asymptotically equal to $\rho$
in the ideal gas limit $\rho\to 0$.

The symmetry of the WR fluid, whose spontaneous breaking is responsible 
for the existence of the liquid and vapor phases, is hidden in the original 
formulation.
It becomes transparent after mapping the WR model in a thermodynamic sense 
onto a two-component WR mixture with non-additive hard-core interactions. 

The WR mixture is composed of two kinds of molecules $\alpha\in\{ A,B\}$ 
interacting pairwisely $U(\{ {\bf r}\}) = 
\sum_{i<j} u_{\alpha\beta}(\vert {\bf r}_i-{\bf r}_j\vert)$,
where the particles of the same species do not interact whereas
the unlike species interact with a hard-core repulsion at distances 
smaller than $2\sigma$, i.e. 
\begin{equation} \label{4}
u_{\alpha\beta}(r) = \left\{
\begin{array}{ll} 
\infty & \mbox{if $\alpha\ne\beta$ and $r<2\sigma$,} \cr
0 & \mbox{otherwise.}
\end{array}
\right.
\end{equation} 
Let $z_A$ and $z_B$ be the dimensionless activities of the two species,
normalized so as to be equal to the densities
$\rho_A = v_0 \langle N_A\rangle/V$ and $\rho_B = v_0 \langle N_B\rangle/V$
in the ideal gas limit, and ${\cal P}_{\rm mix}(z_A,z_B)$ be the dimensionless 
pressure (independent of the temperature since the only interactions are 
infinitely strong repulsions) defined in analogy with (\ref{3}).
The mixture possesses the obvious symmetry with respect to the interchange 
of $A$ and $B$ molecules,
${\cal P}_{\rm mix}(z_A,z_B) = {\cal P}_{\rm mix}(z_B,z_A)$.
Let us consider the line of equivalent activities $z_A=z_B=z$.
The system behaves like the ideal gas at very low $z$,
i.e. there is just one pure (mixed) phase with 
equivalent particle densities $\rho_A=\rho_B$.
At very large $z$, since unlike molecules experience an infinitely
strong repulsion, a mixed phase suffers from packing effects,
which are substantionally reduced in a demixed phase with a single
majority component.
Consequently, the $A-B$ symmetry of the WR mixture is broken and
the system can exist in two different homogeneous (i.e. translation
invariant) pure phases: the $A$-rich phase with $\rho_A-\rho_B=\delta\rho>0$ 
and the complementary $B$-rich phase with $\rho_B-\rho_A=\delta\rho$.
The two phases become equivalent $(\delta\rho=0)$ at the
``demixing'' critical point $z_d$.

The two-component WR mixture is exactly solvable in one dimension
with no phase transition.
For dimensions $d\ge 2$, the proof of the existence of more than one pure
thermodynamic phase for sufficiently large $z$ was given by Ruelle
\cite{Ruelle} using the Peierls contour method.
\medskip

\noindent {\bf Theorem 1.} 
For each dimension $d\ge 2$, there exists $z_d>0$ such that the 
$A$-$B$ WR model on the line of equivalent activities $z_A=z_B=z$ has
\begin{itemize}
\item
A unique homogeneous mixed phase for all $z<z_d$.
\item
Multiple homogeneous demixed phases for all $z>z_d$. 
\end{itemize}

By integrating over the coordinates of one species in the two-component
WR mixture, its thermodynamic equivalence with the original
WR model is realized via the identity between 
the dimensionless pressures \cite{Widom}
\begin{equation} \label{5}
{\cal P}(z,\theta) = - \theta + 
{\cal P}_{\rm mix}(\theta,z{\rm e}^{\theta}) .
\end{equation} 
The line $z_A=z_B$ of the $A$-$B$ mixture is thus transcribed
to the line of symmetry 
\begin{equation} \label{6}
z = \theta \exp(-\theta) 
\end{equation}
in the original WR model and the demixing critical point $z_d$
has its image at the critical point $(\theta^*=z_d,z^*=z_d {\rm e}^{-z_d})$.
For $\theta>\theta^*$, the line of symmetry (\ref{6}) is the coexistence
curve of the gas (vapor) phase with particle density $\rho_g$ and
of the liquid phase with density $\rho_l>\rho_g$.
The mapping enables one to deduce, without solving explicitly the WR model, 
singular behaviors of thermodynamic quantities on the line of symmetry 
(\ref{6}) in the neighborhood of the demixing critical point.
Some features of singular behaviors differ from those predicted by
the lattice gas model \cite{Lee}, e.g. the temperature derivative
of the mean density $(\rho_g+\rho_l)/2$ is proportional to the heat
capacity at constant volume and goes to $-\infty$ at the critical point
\cite{Widom}.

The multicomponent generalization of the $A$-$B$ WR mixture 
consists in considering molecules of $M$ different types 
$\alpha=1,2,\ldots,M$, all having the same activity $z$.
The molecules interact pairwisely, the only interaction is a hard-sphere
repulsion between any two particles of unlike species (\ref{4}).

It was shown \cite{Runnels} that for any finite number $M$ of components
the WR model in $d\ge 2$ dimensions exhibits the demixing phase transition
at some $z_d(M)$ (Theorem 1); in a pure demixed phase, 
the homogeneous density of just one of the components is dominant, 
say $\rho_1>\rho_2=\rho_3=\ldots=\rho_M$.

A version of the $M$-component WR model, in which the hard-sphere 
interaction between unlike species is slightly modified to that 
of parallel hard (hyper-)cubes, was studied in the limit of infinite 
dimensionality $d\to\infty$ \cite{Sear}.
In that limit, the calculation of thermodynamic functions within 
the second virial coefficient is exact.
It turns out that for $M\ge 31$ the transition from the mixed phase
at small values of $z$ to the demixed phase at large values of $z$
is preempted by solidification at intermediate values of $z$, 
$z_c(M) < z < z_d(M)$. 
In the corresponding crystal phase, all species are equivalent 
($\rho_1=\rho_2=\ldots =\rho_M=\rho/M$ with $\rho$ being the total
density of molecules), but the density $\rho\equiv \rho({\bf r})$ 
varies periodically in space, i.e. the translation symmetry is broken. 
The origin of this phenomenon is purely entropic: for a large number
of component $M$ it pays the system to create a periodic structure
of alternating dense and sparse regions, where the particles in 
the dense regions are less restricted by the hard-core repulsions 
coming from particles in the sparse regions.

\medskip 

\noindent {\bf Lattice Widom-Rowlinson model:}
The lattice version of the two-species WR model was introduced by
Lebowitz and Gallavotti \cite{Lebowitz71}.
The model is defined on a regular $d$-dimensional integer lattice
${\rm Z}^d$, each lattice site $i$ can be either empty 
\{local state $\sigma_i=0$, the corresponding activity $z(0)=1$\}
or singly occupied by a particle of type $A$ $\{\sigma_i=1, z(1)=z\}$
or $B$ $\{\sigma_i=2, z(2)=z\}$.
The only interaction is an infinite repulsion between unlike particles
on nearest-neighbor pairs of lattice sites, i.e. the potential energy 
associated with a given state configuration $\sigma$ of all sites is formally 
expressible as $U(\sigma) = \sum_{\langle i,j\rangle} u(\sigma_i,\sigma_j)$,
where the interaction potential between nearest-neighbor pair of lattice
sites $\langle i,j\rangle$ is given by
\begin{equation} \label{7}
u(\sigma_i,\sigma_j) = \left\{
\begin{array}{ll}
\infty & \mbox{if $\sigma_i\ne \sigma_j$ and $\sigma_i\ne 0, \sigma_j\ne 0$,}
\cr
0 & \mbox{otherwise.}
\end{array}
\right.
\end{equation}
A relatively simple proof of the demixing phase transition (Theorem 1)
in the lattice WR model was done using standard Peierls methods 
\cite{Lebowitz71}.
Moreover, in dimensions $d\ge 3$, the existence of a sharp interface 
between coexisting demixed $A$-rich and $B$-rich phases, generated via
the respective fixed $A$ and $B$ boundary conditions on opposite
sides of a box, was proved in the limit of the box size going 
to infinity \cite{Bricmont}.  

In the multicomponent lattice generalization of the two-component WR model,
each lattice site $i\in {\rm Z}^d$ can be either empty
$\{ \sigma_i=0,z(0)=1\}$ or singly occupied by a particle of type
$\sigma_i=1,2,\ldots,M$, all having the same activity $z$.
The potential energy of a state configuration $\sigma$ reads
$U(\sigma) = \sum_{\langle i,j\rangle} u(\sigma_i,\sigma_j)$,
where the interaction potential between nearest-neighbor sites
$\langle i,j\rangle$ is given by relation (\ref{7}).
Note that if one replaces $\infty$ in (\ref{7}) by some finite $U\ne 0$,
the system is equivalent to a dilute Potts model.
The number density of species $\sigma=1,\ldots,M$ at site $i$
will be denoted by $\rho_i(\sigma)$, the total density of
particles at site $i$ by $\rho_i=\sum_{\sigma=1}^M \rho_i(\sigma)$.

The multicomponent lattice WR model was studied by
Runnels and Lebowitz \cite{Runnels}. 
In dimensions $d\ge 2$ and for any finite number $M$ of components,
the lattice WR model exhibits the demixing phase transition at some $z_d(M)$ 
(Theorem 1); in a pure demixed phase, the site-independent density of just 
one of the components is dominant, say $\rho(1)>\rho(2)=\ldots=\rho(M)$.
When the number of components $M$ is larger than some minimum $M_0$,
an entropy-driven crystal phase prevents from a direct transition 
between the mixed and demixed phases.
\medskip

\noindent {\bf Theorem 2.}
For each dimension $d\ge 2$, if the number $M$ of components of the 
lattice WR model exceeds some $M_0$, there exist two activity thresholds,
the crystal one $z_c(M)>0$ and the demixed one $z_d(M)>z_c(M)$, such that:
\begin{itemize}
\item
For $z<z_c(M)$, there exists a unique translation invariant mixed phase 
with species-independent densities $\rho_i(\sigma) = \rho/M$.
\item
For intermediate values of $z$, $z_c(M)<z<z_d(M)$, there exist two
distinct crystalline phases in which one of the alternating sublattices
(even or odd) is preferentially occupied by particles, i.e.
the translation symmetry (but not the species symmetry) is broken.
More precisely, the average particle densities on the even and odd
sublattices are unequal, $\rho_e\ne \rho_o$, while the average densities
of the species $\sigma=1,2,\ldots,M$ are the same within a given sublattice,
i.e. $\rho_e(\sigma)=\rho_e/M$ an $\rho_o(\sigma)=\rho_o/M$.
\item
For $z>z_d(M)$, there exist $M$ translation invariant demixed phases
with the density dominance of just one of the species,
say $\rho(1)>\rho(2)=\ldots=\rho(M)$.
\end{itemize}
The transition at $z_c(M)$ is of second order.
For asymptotically large $M$ the crystal threshold $z_c(M)$ 
behaves like $\lambda_{\rm cr}/M$, where $\lambda_{\rm cr}$ 
is the critical activity of the one-component lattice gas 
with nearest-neighbor hard-core exclusions.
The rigorous upper bound $M_0<27^6$ derived for the square lattice 
\cite{Runnels} was surprisingly large and made the crystallization
phenomenon only of academic interest.  

The exact solution of the multicomponent WR model on 
the Bethe lattice of coordination number $q$ gives 
$M_0=[q/(q-2)]^2$ \cite{Lebowitz95}, which would suggest more realistic 
$M_0\sim 4$ for the square lattice with $q=4$.
The Monte-Carlo (MC) simulations for the square lattice 
\cite{Lebowitz95} imply $M_0=7$.
The MC simulations indicate that the transition at $z_d(M)$ is of second order
for $M\le 4$ and of first order for $M\ge 5$, putting the $M$-component
lattice WR model in the $M$-state Potts model universality class.
For asymptotically large $M$, $z_d(M)\sim M-2+1/M+\cdots$.

The consideration of hard-core exclusion to next-to-nearest-neighbors 
leads to analogous phases which numbers and characters depend on the
specific geometry of the model \cite{Georgii}.


\medskip

\begin{thebibliography}{9}

\bibitem{Lee} Lee TD and Yang CN (1952)  
Statistical Theory of Equations of State and Phase Transitions
II. Lattice Gas and Ising Model.
Physical Review 87(3):410-419

\bibitem{Widom} Widom B and Rowlinson JS (1970)
New Model for the Study of Liquid-Vapor Phase Transitions.
Journal of Chemical Physics 52(4):1670-1684

\bibitem{Ruelle} Ruelle D (1971)
Existence of a Phase Transition in a Continuous Classical System.
Physical Review Letters 27(16):1040-1041

\bibitem{Runnels} Runnels LK and Lebowitz JL (1974)
Phase Transitions of a Multicomponent Widom-Rowlinson Model.
Journal of Mathematical Physics 15(10):1712-1717

\bibitem{Sear} Sear RP (1996)
Ordering in Many Component Widom-Rowlinson models.
Journal of Chemical Physics 104(24):9948-9955

\bibitem{Lebowitz71} Lebowitz JL and Gallavotti G (1971)
Phase Transitions in Binary Lattice Systems.
Journal of Mathematical Physics 12(7):1129-1133

\bibitem{Bricmont} Bricmont J, Lebowitz JL, Pfister ChE, and Olivieri E (1979)
Non-Translation Invariant Gibbs States with Coexisting Phases. 
Communications in Mathematical Physics 66(1):1-20

\bibitem{Lebowitz95} Lebowitz JL, Mazel A, Nielaba P, and {\v S}amaj L (1995)
Ordering and Demixing Transitions in Multicomponent Widom-Rowlinson Models.
Physical Review E 52(6):5985-5996

\bibitem{Georgii} Georgii H-O and Zagrebnov V (2001)
Entropy-Driven Phase Transitions in Multitype Lattice Gas Models.
Journal of Statistical Physics 102(1/2):35-67


\end{thebibliography}
\end{document}